\documentclass[sigconf]{acmart}
\usepackage{subfig}
\usepackage{multirow}
\usepackage{graphicx}
\usepackage{enumitem}
\usepackage{balance}
\usepackage{algorithm}
\usepackage{algorithmic}
\usepackage{hyperref}
\newcommand{\proposed}{\textsf{DAWN}}
\newcommand\HH{
  \global\let\savedtextbullet\textbullet
  \gdef\textbullet{%
    \par\noindent\savedtextbullet\global\let\textbullet\savedtextbullet
  }%
}
\AtBeginDocument{%
  }

\copyrightyear{2025}
\acmYear{2025}
\setcopyright{acmlicensed}
\acmConference[WSDM '25] {Proceedings of the Eighteenth ACM International Conference on Web Search and Data Mining}{March 10--14, 2025}{Hannover, Germany.}
\acmBooktitle{Proceedings of the Eighteenth ACM International Conference on Web Search and Data Mining (WSDM '25), March 10--14, 2025, Hannover, Germany}
\acmISBN{979-8-4007-1329-3/25/03}
\acmDOI{10.1145/3701551.3703524}

\begin{document}

\title{{Revisiting Fake News Detection: Towards Temporality-aware Evaluation by Leveraging Engagement Earliness}}


\author{Junghoon Kim}
\authornote{Both authors contributed equally to this research.}
\affiliation{%
  \institution{KAIST}
  \city{Daejeon}
  \country{Republic of Korea}
  }
\email{jhkim611@kaist.ac.kr}

\author{Junmo Lee}
\authornotemark[1]
\affiliation{%
  \institution{KAIST}
  \city{Daejeon}
  \country{Republic of Korea}}
\email{bubblego0217@kaist.ac.kr}

\author{Yeonjun In}
\affiliation{%
  \institution{KAIST}
  \city{Daejeon}
  \country{Republic of Korea}}
\email{yeonjun.in@kaist.ac.kr}

\author{Kanghoon Yoon}
\affiliation{%
  \institution{KAIST}
  \city{Daejeon}
  \country{Republic of Korea}}
\email{ykhoon08@kaist.ac.kr}

\author{Chanyoung Park}
\authornote{Corresponding author.}
\affiliation{%
  \institution{KAIST}
  \city{Daejeon}
  \country{Republic of Korea}}
\email{cy.park@kaist.ac.kr}


\begin{abstract}
Social graph-based fake news detection aims to identify news articles containing false information by utilizing social contexts, e.g., user information, tweets and comments. However, conventional methods are evaluated under less realistic scenarios, where the model has access to future knowledge on article-related and context-related data during training. In this work, we newly formalize a more realistic evaluation scheme that mimics real-world scenarios, where the data is \textit{temporality-aware} and the detection model can only be trained on data collected up to a certain point in time. We show that the discriminative capabilities of conventional methods decrease sharply under this new setting, and further propose~\proposed, a method more applicable to such scenarios. Our empirical findings indicate that later engagements (e.g., consuming or reposting news) contribute more to noisy edges that link real news-fake news pairs in the social graph. Motivated by this, we utilize feature representations of engagement earliness to guide an edge weight estimator to suppress the weights of such noisy edges, thereby enhancing the detection performance of \proposed. Through extensive experiments, we demonstrate that \proposed~outperforms existing fake news detection methods under real-world environments. The source code is available \textbf{\href{https://github.com/LeeJunmo/DAWN}{here}}.
\end{abstract}




\maketitle

\vspace{-3ex}
\section{Introduction}
The advent of social media platforms has made it possible for news to travel to a vast amount of people, allowing for greater accessibility and efficiency when obtaining information. However, the propagation of news articles containing false information has also become much easier. The spreading of such fake news has a detrimental effect on societal security and public health, ranging from potentially affecting presidential election results~\cite{election} to inciting distrust and panic during a pandemic~\cite{covid1, covid2}. As such, the field of \textit{fake news detection}, which aims to identify fabricated news articles, has gained increasing attention and importance in recent years~\cite{survey1, survey2}.

\begin{figure}[t]
\centering
\vspace{-2ex}
\subfloat[Comparison between the two settings.]{
\includegraphics[width=0.85\linewidth]{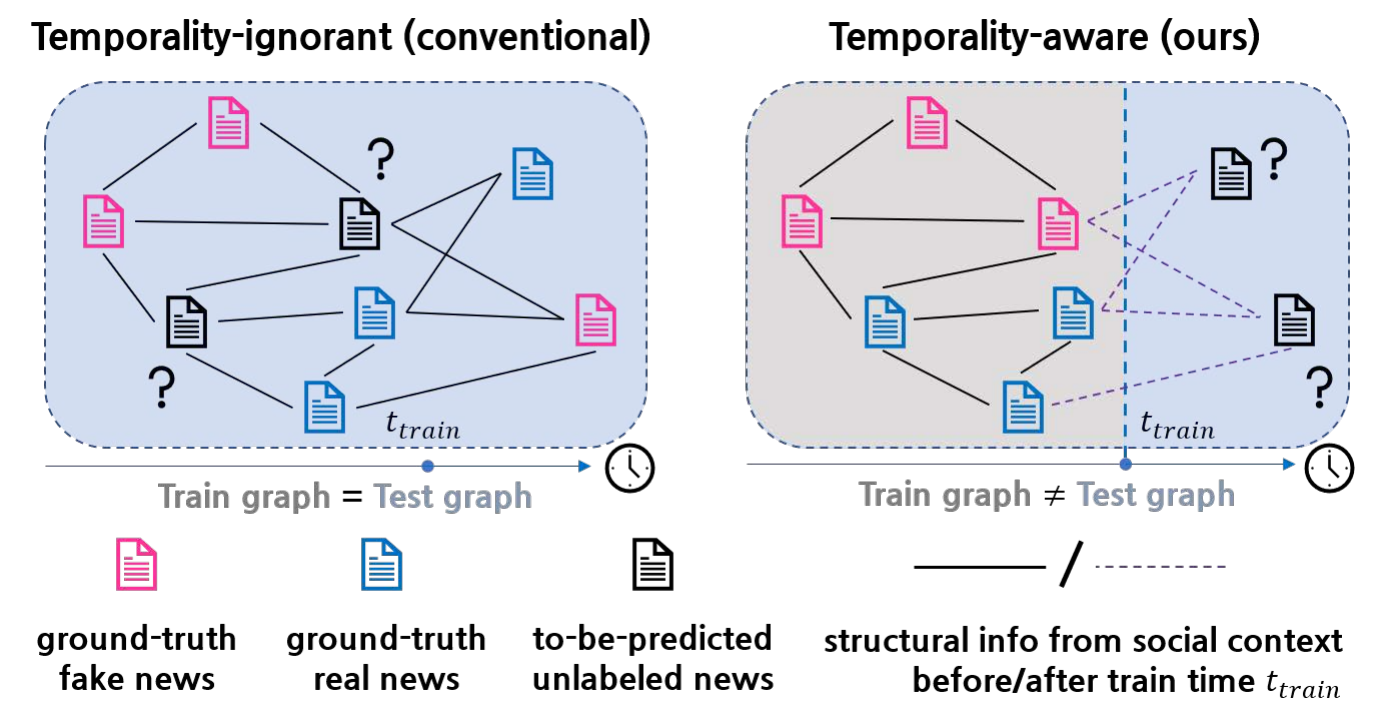}
}
\\
\vspace{-2ex}
\subfloat[Performance of existing methods evaluated under the two settings.]{
\includegraphics[width=0.88\linewidth]{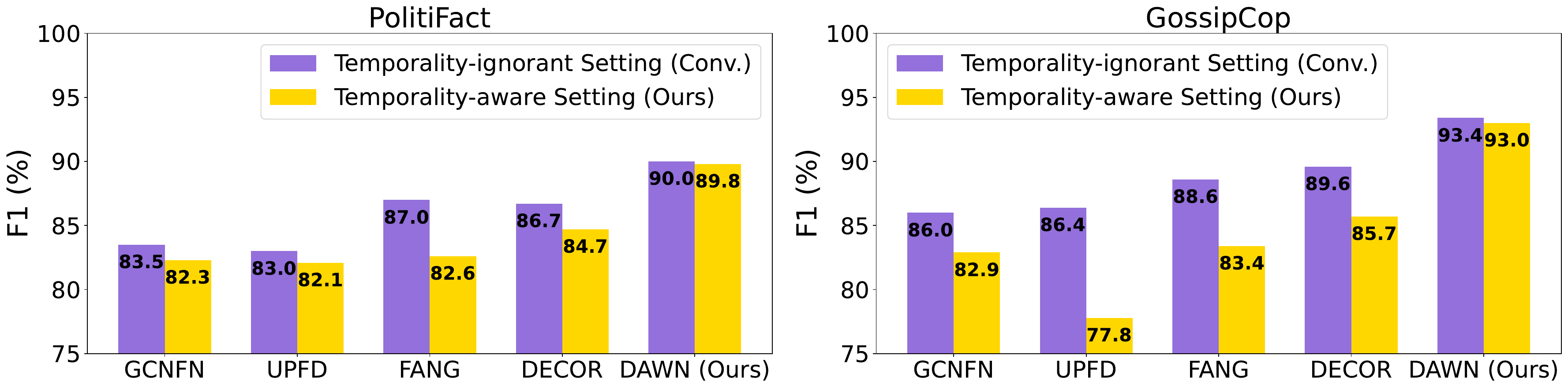}
}
\vspace{-2ex}
\caption{(a) Conventional settings (left) for social graph-based fake news detection utilize future information and social context during training, as opposed to a more realistic temporality-aware setting (right). (b) Existing social graph-based fake news detection methods suffer performance degradation when applied on temporality-aware settings.}
\vspace{-4ex}
\label{fig:intro}
\end{figure}

\looseness=-1
Recent fake news detection methods can be classified into two main categories. \textit{Content-based} methods leverage patterns that can be obtained from the news article text itself, such as semantic representations or emotional features~\cite{bert, style, emotion, ika}. On the other hand, \textit{social graph-based} methods utilize social context knowledge, including user information, tweets and comments, in addition to textual contents~\cite{decor, pna, gcnfn, upfd}. Specifically, social context knowledge can be beneficial for fake news detection since retweets on news pieces can reveal different propagation patterns for real and fake news, while user interactions and responses indicate how these types of news attract different user groups. Existing approaches model such contexts into graph structures, e.g., linking news article pairs based on the amount of tweets shared by readers. By utilizing Graph Neural Networks (GNNs) to model the relationship between news veracity and the structural patterns involved in rich social information, social graph-based methods often outperform their content-based counterparts and appear to be a promising area of research~\cite{decor, defend}.

However, we point out that conventional social graph-based methods are trained and evaluated under an unrealistic scenario~\cite{upfd, decor}. Specifically, they divide news articles into training and test sets via \textit{random split} or use social contexts that occur \textit{after} the training time for training the model. Such \textit{temporality-ignorant} settings (See Fig.~\ref{fig:intro}(a) left) would lead to information leakage as the model can access future knowledge not only on \textit{article-related data} (e.g., textual contents and veracity labels) but also \textit{context-related data} (e.g., users, tweets and comments). We argue that this is inconsistent with real-world scenarios, where the detection model would be trained on offline data collected in advance, and then tested on online data delivered in real-time. Thus, it is more practical for the model to be trained by only utilizing data up to a certain point in time, while future data should only be available at test time, which we call a \textit{temporality-aware} setting (See Fig.~\ref{fig:intro}(a) right).

In Fig.~\ref{fig:intro}(b), we examine the performance of existing social graph-based fake news detection methods (e.g., GCNFN~\cite{gcnfn}, UPFD~\cite{upfd}, FANG~\cite{fang} and DECOR~\cite{decor}) when they are trained and evaluated under the temporality-aware setting. Our empirical results on two prominent fake news datasets~\cite{fakenewsnet} reveal that when all related data are split into train/test considering the temporality (yellow bar), the performance of conventional methods decrease sharply, with a decline up to 8.6\%p in F1 score. Such performance drops can be attributed to the inherent design flaws of these methods; that is, they construct a graph based on the \textit{complete} information regarding social context given in the dataset regardless of the temporal information, leading to potential information leakage, which results in a single fixed structure for both training and testing (See Fig.~\ref{fig:intro}(a) left). On the other hand, in the temporality-aware setting, the structure would change substantially after training (See Fig.~\ref{fig:intro}(a) right).

\looseness=-1
In this work, we revisit the training and evaluation setting of social graph-based fake news detection methods, and propose a novel method that is applicable to real-world learning environments in which the temporal information should not be overlooked. The main idea is to utilize time-independent patterns in the graph, i.e., patterns representing general social user behaviors that occur similarly even at different points in time, so that the model is less affected by how the graph is constructed. Our method, called \textbf{\underline{D}}etecting fake news via e\textbf{\underline{A}}rliness-guided re\textbf{\underline{W}}eighti\textbf{\underline{N}}g (\textbf{\proposed}), takes into account the earliness-related patterns of users and tweets. These patterns are based upon fundamental and well-explored theory on social behaviors, i.e., the confirmation bias theory~\cite{cbias}, and thus the features extracted from them differ less before and after training, making them more robust against temporality-aware settings.

\looseness=-1
Through an extensive data analysis in Section \ref{analysis}, we find that users on social media who have stronger opinions and thus more rapidly engage (e.g., consume or repost) with news articles are more inclined to be attracted to articles of the same veracity, i.e., either only real news or only fake news. From these observations, we obtain some valuable insights: \textbf{(1) earlier engagements tend to connect articles of the same veracity}, while \textbf{(2) the existence of later engagements leads to a higher probability of the labels being different}, which can manifest as noisy edges in social graphs (i.e., edges linking real news-fake news pairs). Based on our findings suggesting that appropriately utilizing knowledge on \textit{engagement earliness} can help identify the edge noise, we suppress the weights of such noisy edges aiming at enhancing the model's discriminative capabilities. To this end, we employ a Graph Structure Learning (GSL) framework, where we train an edge weight estimator to assign new weights to existing edges. By leveraging feature representations of edge-specific engagement earliness, we can successfully guide the edge weight estimator to downweight noisy edges.

Overall, our contributions can be summarized as follows:
\begin{itemize}[leftmargin = 2mm]
  \item \textbf{A realistic scenario for fake news detection. } We argue that the assumption made by existing social graph-based methods that all relevant information is accessible during training is unrealistic, leading to potential information leakage. As such, we present a temporality-aware setting for fake news detection (both article-wise and context-wise) in which existing methods underperform owing to their inherent design flaws.  
  \item \textbf{Earliness-related empirical findings. } We analyze how the engagement earliness regarding users and tweets in a social network relate to the veracity label consistency of news article pairs.
  \item \textbf{Earliness-guided GSL. } We propose a novel GSL-based fake news detection framework, called \proposed, that leverages our earliness-related insights to downweight noisy edges in a social graph.
  \item \textbf{Effectiveness. } \proposed~outperforms existing methods by a large margin on two real-world fake news datasets, highlighting the robustness of our simple yet effective earliness-based framework under the temporality-aware training and evaluation setting.
\end{itemize}

\vspace{-2ex}
\section{Related Works}
\subsection{Fake News Detection}
Fake news detection aims to determine whether a given news article contains false information or not, and can generally be seen as a binary classification task that predicts its \textit{veracity label} (0 if real, 1 if fake). Among the two main categories for fake news detection research, \textbf{content-based} methods utilize patterns from within the news article text itself. Such patterns include semantic representations~\cite{bert, roberta}, emotional features~\cite{emotion, ika} and writing style~\cite{style, ika}.

Meanwhile, \textbf{social graph-based} methods~\cite{gcnfn, decor, fang, pna, upfd} additionally incorporate various social contexts including user information, tweets and comments, and have shown state-of-the-art performance, generally improving over content-based methods. For instance, GCNFN~\cite{gcnfn} constructs propagation trees for each news article by utilizing user responses and user following relations. FANG~\cite{fang} learns the representations of a heterogeneous social graph, consisting of users, news articles and sources. DECOR~\cite{decor} links news article pairs based on co-user engagement knowledge. The social graphs constructed in these works utilize the \textit{entire} information on social context, resulting in a single fixed structure for both training and testing. In other words, they are inherently designed to leverage future contextual data for model training leading to potential information leakage, which especially aggravates when the news articles are randomly split as well.

In this paper, we emphasize the importance of simulating real-world learning environments to enhance the applicability of fake news detection methods. Specifically, as information that appears during test time would not be available at training time under real-world scenarios, such \textit{temporality-aware} settings should be properly deployed when evaluating model performance. To the best of our knowledge, we are the first to consider \textbf{\textit{both} article-wise (textual contents and veracity labels) and context-wise (social knowledge on users and tweets) temporality-aware settings} for social graph-based fake news detection.

\subsection{Graph Structure Learning}
Various downstream tasks on graphs have been shown to be successfully tackled by Graph Neural Networks (GNNs)~\cite{gcn, gat, graphsage}. However, the performance of such GNNs are vulnerable to the existence of \textit{noisy edges} that connect dissimilar nodes~\cite{noisyfeature}, either through structural adversarial attacks or inherent noise within the data~\cite{attack1, attack2, rsgnn, attack3, ours1, ours3, ours4}. As such, many Graph Structure Learning (GSL) studies have aimed to downweight such noisy edges by optimizing the adjacency matrix. Guided by node feature similarity, prior studies have utilized various similarity metrics and link predictor training schemes to alleviate the effect of edge noise~\cite{noisyfeature, jaccard, prognn, ours2, rsgnn}.

A recent study, called DECOR~\cite{decor}, has attempted to apply GSL to fake news detection, and has shown that constructing edge-specific features representing social patterns is much more effective than previous node feature-based methods when training the link predictor to suppress noisy edges. Despite its effectiveness, the degree-related patterns utilized by DECOR are obtained from a single fixed social graph, i.e., the patterns are subject to substantial changes under real-world environments where the graph structure greatly differs before and after training time, hindering detection performance. Motivated by this, our work aims to exploit new features whose underlying patterns are independent of time and thus are more robust under temporality-aware settings.

\section{Preliminaries}
\subsection{Problem Statement}
\noindent{\textbf{Definitions. }}
Let $\mathcal{D = (P, U, R)}$ be a fake news detection dataset. $\mathcal{P}$ is a set of questionable news articles, where each news article $p_n \in \mathcal{P}$ contains the corresponding article text and the time of its publication $t_n ^p$. $\mathcal{U}$ is a set of active users on social media, where each user has had at least $m$ engagements (an \textit{engagement} occurs when a user reposts a news article). $\mathcal{R}$ is a set of such engagements, where each engagement $r_l \in \mathcal{R}$ is defined as $\{(u, p, t_l ^r) | u \in \mathcal{U}, p \in \mathcal{P}\}$ (i.e., user $u$ has reposted news article $p$ at time $t_l ^r$).

To properly emulate real-world scenarios where accessible information changes by time, we adopt a \textit{temporality-aware} setting. Specifically, we define timestamps $t_{train}, t_{val}$ and $t_{test}$ according to how much of the data we want to include in the training, validation and test set, respectively ($t_{train} < t_{val} < t_{test}$).
Then, we construct subsets from $\mathcal{P, U}$ and $\mathcal{R}$. More precisely, from $\mathcal{P}$, we obtain $\mathcal{P}_{train} = \{p_n | p_n \in \mathcal{P}, t_n ^p \le t_{train}\}, \mathcal{P}_{val} = \{p_n | p_n \in \mathcal{P}, t_{train} < t_n ^p \le t_{val}\}$ and $\mathcal{P}_{test} = \{p_n | p_n \in \mathcal{P}, t_{val} < t_n ^p \le t_{test}\}$. Similarly, $\mathcal{R}_{train} = \{r_l | r_l \in \mathcal{R}, t_l ^r \le t_{train}\}, \mathcal{R}_{val} = \{r_l | r_l \in \mathcal{R}, t_l ^r \le t_{val}\}$ and $\mathcal{R}_{test} = \{r_l | r_l \in \mathcal{R}, t_l ^r \le t_{test}\}$. Accordingly, $\mathcal{U}_{train}, \mathcal{U}_{val}$ and $\mathcal{U}_{test}$ are defined so that they contain users who have at least $m$ engagements in $\mathcal{R}_{train}, \mathcal{R}_{val}$ and $\mathcal{R}_{test}$, respectively.
Finally, $\mathcal{Y}_{train}$ and $\mathcal{Y}_{val}$ contain ground-truth veracity labels (1 if fake. 0 otherwise) associated with news articles in $\mathcal{P}_{train}$ and $\mathcal{P}_{val}$, respectively.

\noindent{\textbf{Temporality-aware Fake News Detection. }}
Given a news dataset $\mathcal{D = (P, U, R)}$ and timestamps $t_{train}, t_{val}$ and $t_{test}$, our goal is to learn a fake news detector, i.e., binary classifier. Specifically, we first train the classifier on $(\mathcal{P}_{train}, \mathcal{U}_{train}, \mathcal{R}_{train})$ and validate on $(\mathcal{P}_{val}, \mathcal{U}_{val}, \mathcal{R}_{val})$. Then, given $(\mathcal{P}_{test}, \mathcal{U}_{test}, \mathcal{R}_{test})$, the classifier would predict the veracity labels $\mathcal{Y}_{test}$ for news articles in $\mathcal{P}_{test}$.

\vspace{-2ex}
\subsection{Social Graph Construction} \label{construction}
Following \cite{decor, pna}, we construct a social graph that effectively captures the relationship between news articles through user engagements. Specifically, we first obtain an \textit{engagement matrix} $E \in \mathbb{R}^{|\mathcal{U}| \times |\mathcal{P}|}$, where each element $E_{ij}$ represents the number of times user $u_i$ has interacted with news article $p_j$, the value of which being the number of the associated engagements in $\mathcal{R}$. Then, we construct a graph $\mathcal{G} = (\mathcal{P, A})$, where a node denotes an article and an edge denotes the co-engagement patterns between a pair of articles. Specifically, the associated adjacency matrix $\mathcal{A} \in \mathbb{R}^{|\mathcal{P}| \times |\mathcal{P}|}$ is generated by retrieving co-engagement patterns from the engagement matrix, i.e., $\mathcal{A} = E^{\top}E$. Each element $\mathcal{A}_{ij}$ represents the edge weight between a pair of news articles $p_i$ and $p_j$, where a weight of zero indicates no shared users between them (i.e., none of the users have reposted both articles). If there are multiple users who have reposted multiple responses on both articles, the weight would increase accordingly. In other words, the edge weights in $\mathcal{A}$ denote the intensity of the co-engagement between article pairs.

To incorporate the realistic temporality-aware setting, we expand on the above procedure and construct three separate graphs for training, validation and testing. In detail, we obtain a training engagement matrix $E_{train} \in \mathbb{R}^{|\mathcal{U}_{train}| \times |\mathcal{P}_{train}|}$ only utilizing $(\mathcal{P}_{train}, \mathcal{U}_{train}, \mathcal{R}_{train})$. From this we construct a training graph $\mathcal{G}_{train} = (\mathcal{P}_{train}, \mathcal{A}_{train})$, where $\mathcal{A}_{train} \in \mathbb{R}^{|\mathcal{P}_{train}| \times |\mathcal{P}_{train}|}$ is defined as $\mathcal{A}_{train} = {E_{train}}^{\top}E_{train}$. Similarly, we construct validation and test graphs $\mathcal{G}_{val} = (\mathcal{P}_{val}, \mathcal{A}_{val})$ and $\mathcal{G}_{test} = (\mathcal{P}_{test}, \mathcal{A}_{test})$ from $(\mathcal{P}_{val}, \mathcal{U}_{val}, \mathcal{R}_{val})$ and $(\mathcal{P}_{test}, \mathcal{U}_{test}$, $\mathcal{R}_{test})$, respectively.

\vspace{-2ex}
\section{Data Analysis: Engagement Earliness and Veracity Label Consistency} \label{analysis}
In this section, we explore the relationship between engagement earliness and the veracity label consistency of news article pairs. Our analysis is performed on two prominent fake news datasets PolitiFact and GossipCop from the FakeNewsNet~\cite{fakenewsnet} benchmark.

\begin{figure}[t]
\centering
\subfloat{
\includegraphics[width=0.4\linewidth]{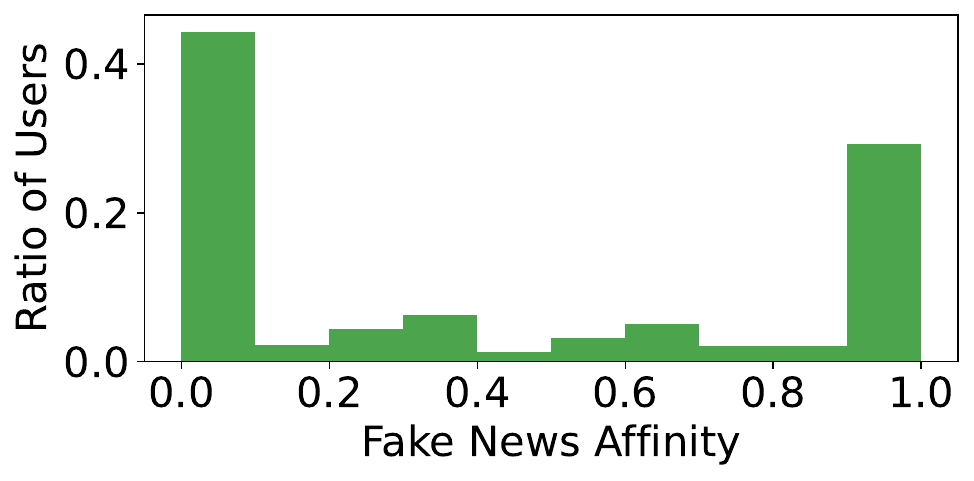}
}
\subfloat{
\includegraphics[width=0.4\linewidth]{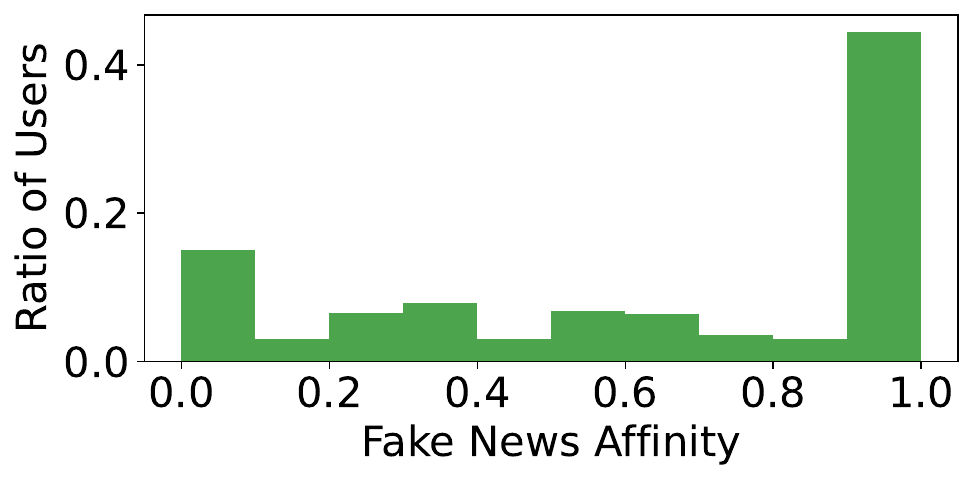}
}
\vspace{-2ex}
\caption{Users have a tendency to engage with either only fake news or real news (Left: PolitiFact, right: GossipCop).}
\vspace{-5ex}
\label{fig:orig-plots}
\end{figure}

Following \cite{pna}, we define a Fake News Affinity (FNA) score for each user $u \in \mathcal{U}$ (we set $m=3$ for all analysis in this section) as:

\begin{equation}
\mathit{FNA(u)} = \frac{\text{\# of engagements with fake news by } u}{\text{\# of all engagements by } u}.
\end{equation}
Fig.~\ref{fig:orig-plots} shows the distribution of users' FNA scores via histograms. The results indicate that users generally have FNA scores close to 1 (i.e., only engage with fake news) or 0 (i.e., only engage with real news), implying that users tend to consume and spread news articles of the same veracity. 
In other words, users engaging with news articles on social media would be attracted to similar articles regarding veracity, according to their highly polarized opinions~\cite{polar1, polar2}.
This is in line with the well-known confirmation bias theory~\cite{cbias} stating that people have a tendency to be drawn to information that affirms and reinforces their prior beliefs and preferences.

Building upon this, we now investigate how the behaviors change in terms of \textit{engagement earliness}. More precisely, as users with stronger opinions would generally act quicker when consuming and reposting news (i.e., engagements) that reflect their beliefs, we  hypothesized that \textit{confirmation bias would exacerbate in users displaying earlier engagement patterns}. In the following, we investigate earliness patterns in terms of each engagement in $\mathcal{R}$ and user in $\mathcal{U}$.

\subsection{Engagement-wise Earliness Patterns}
First, we defined a deadline $t_d$, where a certain engagement $r_l \in \mathcal{R}$ can be considered \textit{early} if $t_r ^l - t_n ^p < t_d$, meaning the engagement occurred within a fixed timespan determined by the deadline after the corresponding news article has been posted. Engagements that occur after the deadline can be considered as \textit{late}. We divided all engagements into two groups (early and late) according to the deadline (set to 30 minutes for PolitiFact and 5 minutes for GossipCop), then obtained the FNA scores separately within each group. 

\begin{figure}[t]
\centering
\subfloat[PolitiFact]{
\includegraphics[width=0.8\linewidth]{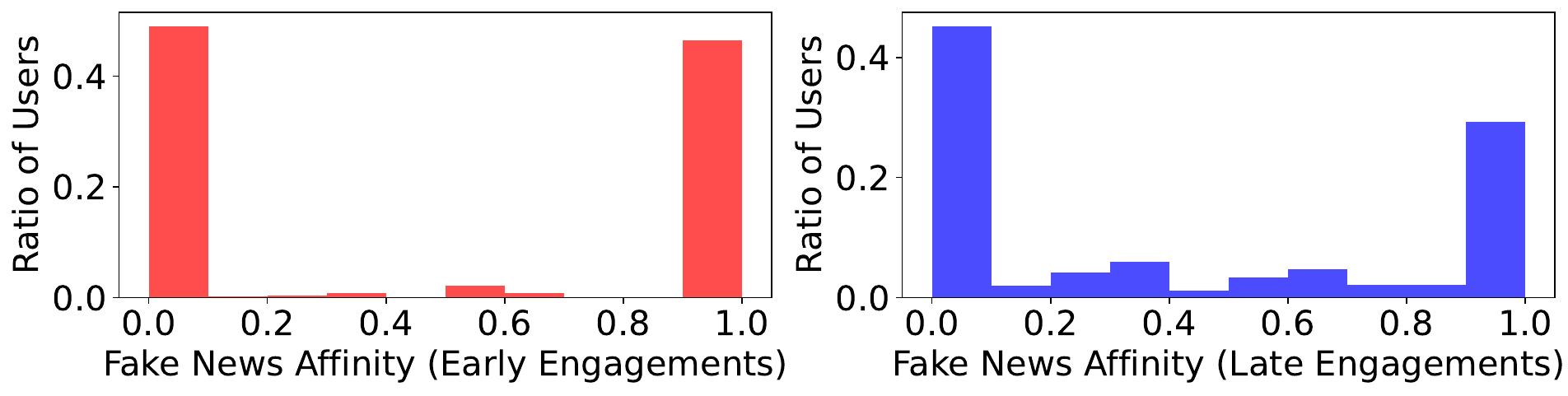}
}
\vspace{-2ex}
\\
\subfloat[GossipCop]{
\includegraphics[width=0.8\linewidth]{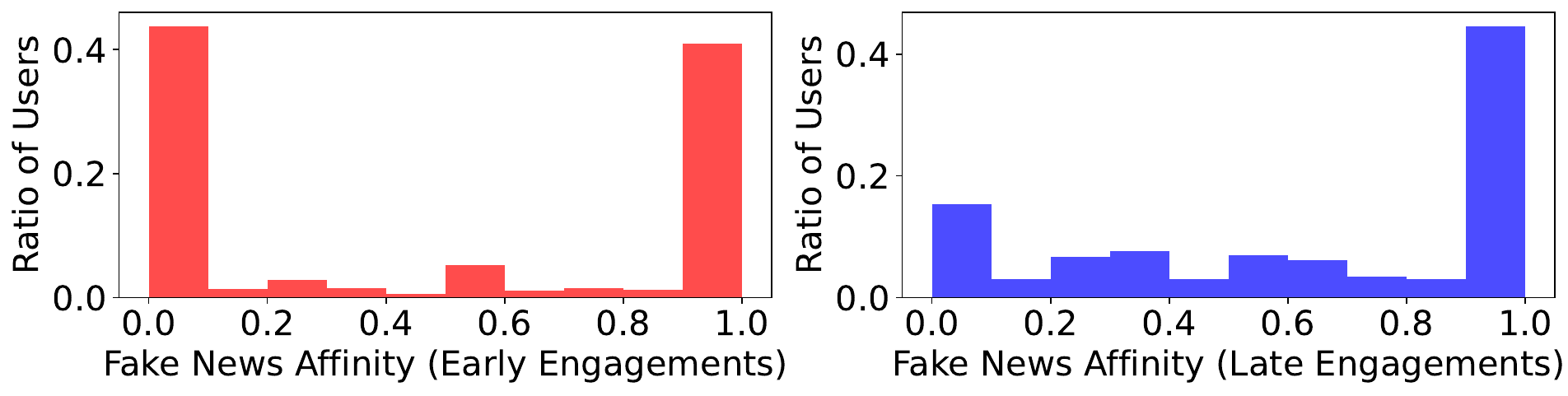}
}
\vspace{-2ex}
\caption{The skewed tendency intensifies with early engagements (red) over late engagements (blue).}
\vspace{-5ex}
\label{fig:eng-plots}
\end{figure}

\begin{figure}[t]
\centering
\subfloat[PolitiFact]{
\includegraphics[width=0.8\linewidth]{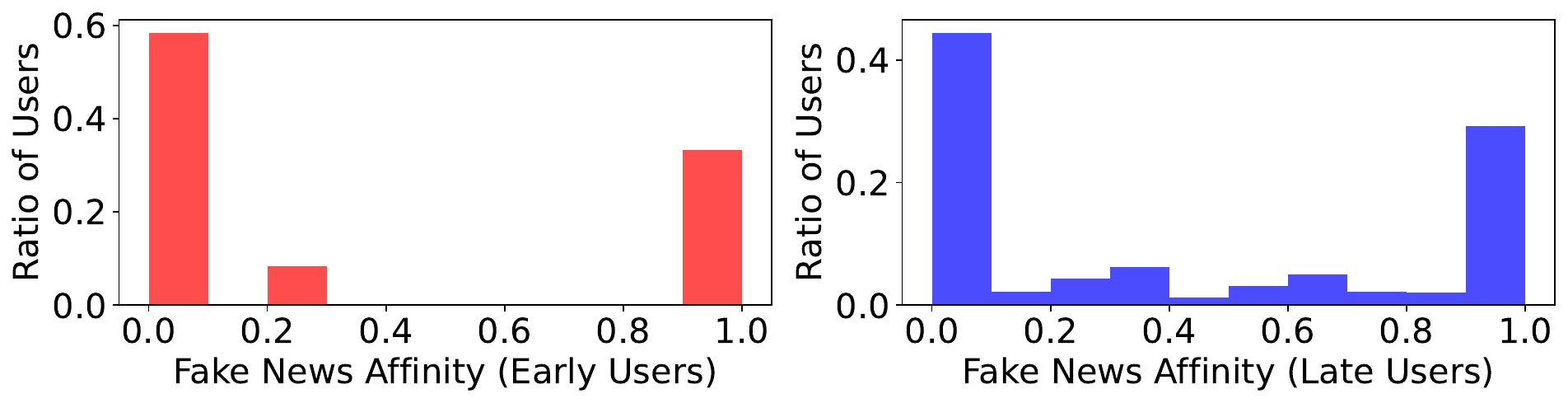}
}
\vspace{-2ex}
\\
\subfloat[GossipCop]{
\includegraphics[width=0.8\linewidth]{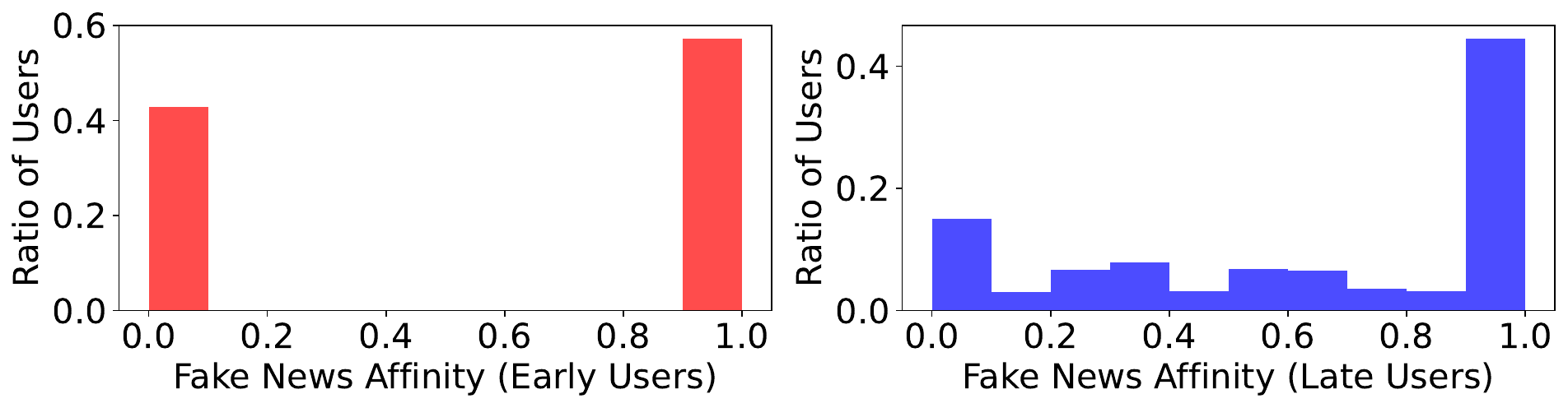}
}
\vspace{-2ex}
\caption{The skewed tendency intensifies with engagements by early users (red) over late users (blue).}
\vspace{-3ex}
\label{fig:us-plots}
\end{figure}

\begin{figure}[t]
\centering
\subfloat[PolitiFact]{
\includegraphics[width=0.8\linewidth]{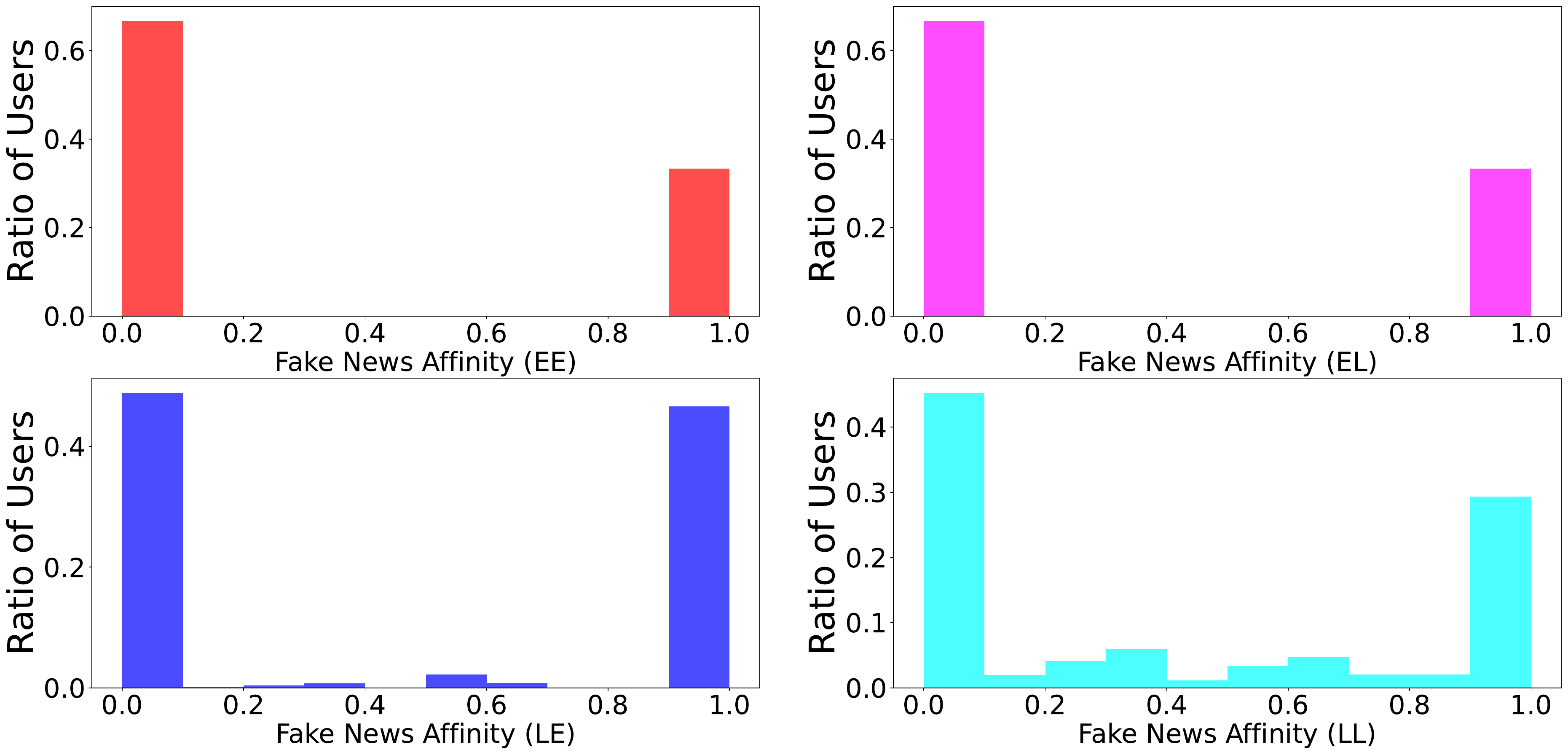}
}
\vspace{-2ex}
\\
\subfloat[GossipCop]{
\includegraphics[width=0.8\linewidth]{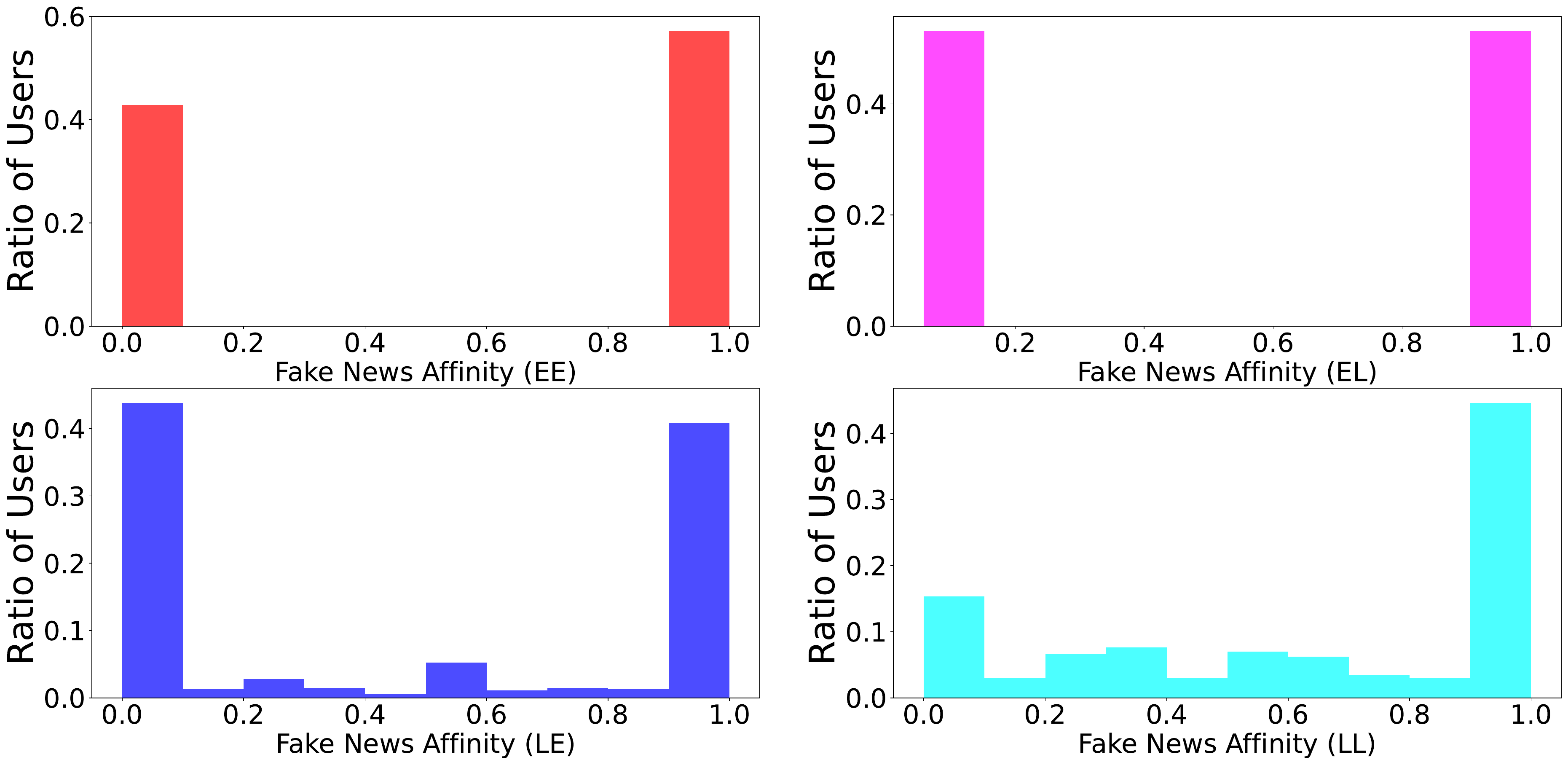}
}
\vspace{-2ex}
\caption{The patterns can be further broken down when considering \textit{both} engagement-wise and user-wise viewpoints.}
\vspace{-3ex}
\label{fig:joint-plots}
\end{figure}

The difference in distributions are shown in Fig.~\ref{fig:eng-plots}, where the FNA scores associated with early engagements are much more skewed towards 0 or 1, compared to those associated with late engagements. In other words, we can observe that confirmation bias does indeed occur more intensely in users with ``earlier engagements.''

\subsection{User-wise Earliness Patterns}
While our previous analysis assigned earliness categories in terms of each engagement, we also observed patterns when they are assigned in terms of each user. Specifically, we defined a User Earliness (UE) score for each user as:
\begin{equation}
\mathit{UE(u)} = \frac{\text{\# of early engagements by } u}{\text{\# of all engagements by } u},
\end{equation}
where early engagements are determined by the previously defined deadline. Users whose UE score exceeds a certain threshold $thres_u$ are classified as \textit{early users}; those that don't are \textit{late users}.

The distributions of the FNA scores of the two different user groups ($thres_u$ was set to 0.8 for both datasets) are shown in Fig.~\ref{fig:us-plots}. We can once again observe that the skewness in the FNA scores increases among early users as opposed to that among late users. Such results indicate that engagements by ``earlier users'' are more likely to display patterns of confirmation bias.

\smallskip
\subsection{Joint Earliness Patterns} \label{Joint Patterns}
Finally, we explore what happens when the previous viewpoints - both engagement-wise and user-wise - are taken into account simultaneously. Keeping $t_d$ and $thres_u$ the same as before, we divided all engagements into four groups: early users' early engagements (EE), early users' late engagements (EL), late users' early engagements (LE) and late users' late engagements (LL).

The distributions of the FNA scores are shown in Fig.~\ref{fig:joint-plots}, where we observe some interesting results. \textbf{(1)} While in our previous observation the late engagement group's FNA scores were less skewed (See Fig.~\ref{fig:eng-plots}), dividing the group further through user-wise earliness we can see that group EL is more skewed compared to group LL. \textbf{(2)} Similarly, dividing the late user group further through engagement-wise earliness, it can be seen that group LE is more skewed compared to group LL.

\smallskip
\noindent{\textbf{Implications. }}
Our extensive analyses on earliness-related patterns reveal that earlier user engagements have a stronger tendency to be linked with news articles of the same veracity, supporting our hypothesis. The major implication is that within the social graph constructed through the procedure detailed in Section \ref{construction}, \textit{edges containing later engagements have a higher likelihood of connecting ``real news''-``fake news'' pairs than those containing earlier engagements}. In other words, in the original $\mathcal{A}$ where all engagements are treated equally when determining edge weight, such ``later'' edges are more likely to be noisy edges that hinder the performance of existing social graph-based fake news detection models.

\section{Proposed Method:~\proposed} \label{method}
\begin{figure}[t]
\centering
\includegraphics[width=0.9\linewidth]{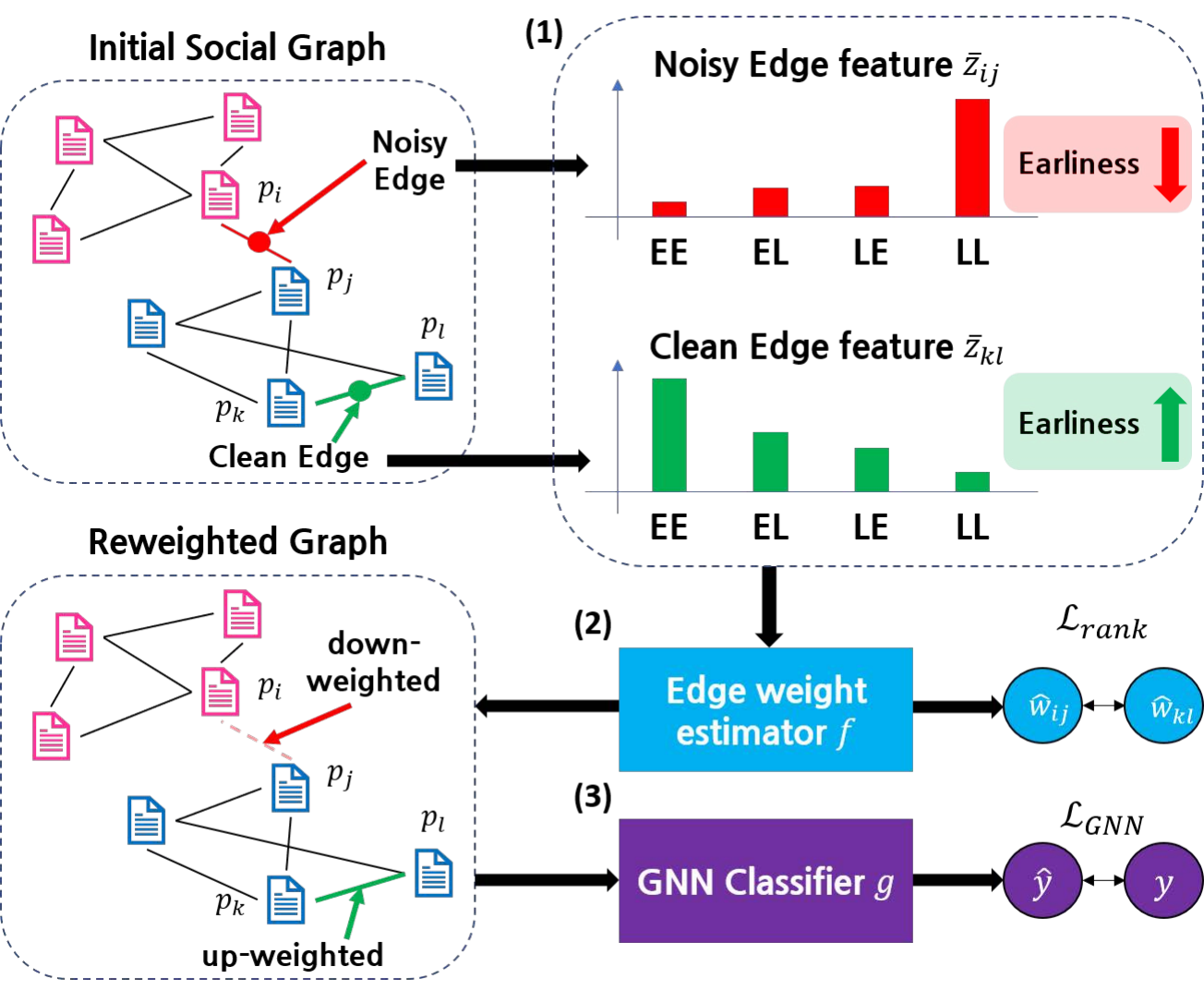}
\caption{Overall framework of \proposed. After (1) extracting edge-specific features representing engagement earliness, (2) the features are fed into an edge weight estimator $f$, which distinguishes existing clean and noisy edges and adjusts their weights guided by a ranking loss. (3) Finally, a GNN classifier $g$ is applied on the reweighted graph to predict node veracity.}
\vspace{-2ex}
\label{fig:architecture}
\end{figure}

Based on our findings regarding engagement earliness, we propose~\textbf{\proposed}, a novel method for \textbf{\underline{D}}etecting fake news via e\textbf{\underline{A}}rliness-guided re\textbf{\underline{W}}eighti\textbf{\underline{N}}g. Fig.~\ref{fig:architecture}
illustrates the overall framework of \proposed, consisting of three components. \textbf{(1)} We first construct edge-specific features for each existing edge in the social graph. Stemming from our insights in Section~\ref{analysis}, we represent the joint earliness patterns within the edges as 4-dimensional vectors. \textbf{(2)} The features are then fed into an edge weight estimator $f$, which is a link predictor aiming to adjust the weights of existing edges. Further guided by a \textit{ranking loss}, $f$ suppresses the weights of noisy edges as opposed to that of clean edges (\textit{noisy edges} connect real news-fake news pairs, while \textit{clean edges} connect real news-real news or fake news-fake news pairs), resulting in a reweighted adjacency matrix $\mathcal{W}$. \textbf{(3)} Finally, a GNN classifier $g$ utilizes the newly obtained $\mathcal{W}$ alongside node features extracted from the article texts to predict the veracity labels of the nodes.

\subsection{Edge Feature Construction}
Our empirical analysis indicates that earliness-related engagement patterns can help in determining the likelihood of a certain edge in the social graph being clean or noisy. Motivated by this, we construct earliness-related features specific to each edge, which can in turn successfully identify ``later'' edges. Our observations regarding the simultaneous consideration of both engagement-wise and user-wise earliness patterns suggest that focusing on either one on its own can lead to critical loss of information, e.g., ignoring the entire late user group will result in ignoring the LE group as well, despite them displaying more skewed patterns.

As such, we design the edge features as joint representations that consider both viewpoints. Let $Z \in \mathbb{R}^{|\text{\# edges}| \times 4}$ denote an edge feature matrix where each row corresponds to a 4-dimensional vector $z_{ij}\in\mathbb{R}^4$, the feature of the edge $\mathcal{A}_{ij}$ connecting news articles $p_i$ and $p_j$. Each $z_{ij}$ is constructed by following the strategy detailed in Section~\ref{Joint Patterns}, where we divide the engagements corresponding to $\mathcal{A}_{ij}$ into four groups, i.e., EE, EL, LE and LL. Specifically, each element of $z_{ij}$ represents the size of each group, respectively. After going through all existing edges, we perform a column-wise normalization on $Z$ to obtain $\bar{Z}$, i.e., $z_{ij}$ is normalized to $\bar{z}_{ij}$.

\subsection{Noisy Edge Suppression Module} \label{reweighter}
The normalized edge features are then fed into an MLP-based edge weight estimator $f$ to obtain adjusted edge weights. This process is formulated as:
\begin{equation}
w_{ij} = f(\bar{z}_{ij}) = sigmoid(MLP(\bar{z}_{ij})),
\label{eqn:mlp}
\end{equation}
where the MLP outputs a single value that is then activated through a sigmoid function, resulting in estimated edge weight $w_{ij} \in [0, 1]$. 
Our objective is to enable $f$ to learn earliness-related patterns in $\bar{z}_{ij}$, allowing it to assign
decreased weights to noisy edges and increased weights to clean edges within a fixed range. By replacing the edge weights of the original adjacency matrix $\mathcal{A}$ with the estimated weights, we can obtain a reweighted adjacency matrix $\mathcal{W}$.

To further guide $f$ in distinguishing noisy edges displaying later patterns from clean edges displaying earlier patterns, we additionally introduce a regularization term for the estimated edge weights. Specifically, as we have access to the ground-truth veracity labels of news articles in the training set, we can divide all edges within the training graph into clean edge or noisy edge groups. A straightforward way of utilizing this information would be through a binary classification loss~\cite{rsgnn}, penalizing noisy and clean edge weights when they are far from 0 and 1, respectively. However, as the rate of earliness is different for each edge, we discover that ignoring this and strictly sending all weights to 0 or 1 hinders performance (refer to Section~\ref{ablation}). As such, we impose constraints that guide reweighting in a less strict manner, by utilizing a ranking loss. Specifically, we randomly sample $K$ edges from both clean and noisy edge groups, and then emphasize the difference within each clean edge-noisy edge pair via the following loss function:

\vspace{-2ex}
\begin{equation}
\mathcal{L}_{rank} = \frac{1}{K^2}\sum_{i=1}^{K}\sum_{j=1}^{K}\max(0, -(w_{clean}^{(i)} - w_{noisy}^{(j)}) + margin),
\label{eqn:margin}
\end{equation}
where $w_{clean}^{(i)}$ and $w_{noisy}^{(j)}$ denote the estimated edge weight of the $i$-th and $j$-th sample from the clean edge and noisy edge group, respectively. By penalizing cases where clean edges have smaller weights than noisy edges, further enhanced by a preset $margin$ value, we can give regularization to relatively reduce the influence of noisy edges. During training, $\mathcal{L}_{rank}$ can guide the edge weight estimator $f$ to obtain a reweighted training adjacency matrix $\mathcal{W}_{train}$. {While comparing \textit{all} possible pairs during training would be ideal, we empirically show that observing $K^2$ sampled pairs still achieves competitive performance in Section~\ref{hyperparameter}.}

\subsection{Fake News Detection Module}
Following prior studies~\cite{decor, upfd}, we extract the initial node feature $x_n$ of a news article $p_n\in \mathcal{P}$ from the article text via a pre-trained BERT~\cite{bert}.
Utilizing the node features and reweighted adjacency matrix $\mathcal{W}$ obtained via edge weight estimator $f$, we can learn the representation of nodes through expressive GNN architectures~\cite{gcn, gat, graphsage}. Based on this learned representation, the veracity of article $p_n$ is predicted in the form of $\hat{y}_n = softmax(h_n)$. $h_n \in \mathbb{R}^2$ is the output of the GNN classifier $g(\mathcal{X, W})$ for article $p_n$, where $\mathcal{X}$ is the collection of all relevant node features in the form of a single matrix. During training, the inputs for $g$ are the node features corresponding to news articles in the training set and $\mathcal{W}_{train}$, and the resulting GNN prediction loss is as follows:

\vspace{-2ex}
\begin{equation}
\mathcal{L}_{GNN} = \sum_{p_n\in \mathcal{P}_{train}}l(\hat{y}_n, y_n),
\end{equation}
$l(\hat{y}_n, y_n)$ denoting the cross entropy between $\hat{y}_n$ and $y_n \in \mathcal{Y}_{train}$.

\subsection{Final Training Objective}
The final loss function for training is as follows:

\vspace{-2ex}
\begin{equation}
\mathcal{L}_{final} = \underset{\theta, \phi}{\arg\min}\mathcal{L}_{GNN} + \alpha \mathcal{L}_{rank},
\label{eqn:final}
\end{equation}
where $\theta$ and $\phi$ are the learnable parameters of GNN classifier $g$ and edge weight estimator $f$, respectively. $\alpha$ is a hyperparameter for balancing the contribution of the ranking loss.~\proposed~follows an end-to-end approach in which $f$ and $g$ are learned simultaneously.

After training on $\mathcal{G}_{train}$, $f$ adjusts the weights of $\mathcal{A}_{val}$, resulting in a reweighted $\mathcal{W}_{val}$. The best performing model on the validation set is then used for final prediction on the test set. Similarly, we adjust the test adjacency matrix through $f$ to obtain $\mathcal{W}_{test}$, which is then fed into $g$ alongside the node features. Through this procedure, \proposed~effectively mitigates the effect of noisy edges and detects fake news using low-dimensional earliness-related edge features. The pseudocode for \proposed~ can be found in Appendix~\ref{pseudocode}.

\section{Experiments}
In this section, we conduct comprehensive experiments to answer the following research questions:
\begin{itemize}[leftmargin = 2mm]
  \item \textbf{RQ1. } How well does our proposed \proposed~ perform in detecting fake news compared with baselines?
  \item \textbf{RQ2. } How effective are our constructed edge features and additional modules in enhancing \proposed's performance?
  \item \textbf{RQ3. } How does \proposed~perform under various hyperparameters?
  \item \textbf{RQ4. } How efficient is \proposed~ on large-scale social networks?
  \item \textbf{RQ5. } Does \proposed~ successfully downweight noisy edges?
\end{itemize}

\subsection{Experimental Setup}
\noindent{\textbf{Datasets. }}
We evaluate \proposed~ on two prominent fake news datasets, i.e., PolitiFact and GossipCop from the FakeNewsNet benchmark~\cite{fakenewsnet}. They contain news articles labeled as real or fake by leading fact-checking websites, along with related tweets by users on X (formerly known as Twitter). The statistics can be found in Table~\ref{table:stats}.

Under our temporality-aware evaluation scheme, $t_{train}, t_{val}$ and $t_{test}$ are set so that 70\%, 10\% and 20\% of the news articles (in temporal order) are assigned to the training, validation and test sets, respectively. The social contexts regarding users and tweets are also split accordingly.

\begin{table}[t]
\centering
\caption{Dataset statistics.}
\vspace{-2ex}
\resizebox{0.8\linewidth}{!}{
\begin{tabular}{l c c}
\hline
\textbf{Dataset}&\textbf{PolitiFact}&\textbf{GossipCop}\\
\hline
\# News Articles&597&8,763\\
\# Real News&282&6,764\\
\# Fake News&315&1,999\\
\# Users&162,262&129,820\\
\# Tweets (Engagements)&255,227&516,172\\
\hline
\end{tabular}
}
\label{table:stats}
\vspace{-3ex}
\end{table}

\smallskip
\noindent{\textbf{Baselines. }}
We compare \proposed~ with the following baselines, which can be further categorized by model architecture: content-based methods (G1) (dEFEND{\textbackslash}c~\cite{defend}, DualEmo{\textbackslash}c~\cite{dualemo}, BERT~\cite{bert} and GPT3.5), social graph-based methods (G2) (GCNFN~\cite{gcnfn}, UPFD~\cite{upfd}, FANG~\cite{fang}, GCN~\cite{gcn}, GAT~\cite{gat} and GraphSAGE~\cite{graphsage}) and Graph Structure Learning methods (G3) (RS-GNN~\cite{rsgnn} and DECOR~\cite{decor}). Further details regarding the baslines are described in Appendix~\ref{exp}.1.

\smallskip
\noindent{\textbf{Implementation Details. }}
The implementation details for our proposed \proposed~ are described in Appendix~\ref{exp}.2.

\smallskip
\noindent{\textbf{Evaluation Metrics. }}
Following previous works~\cite{upfd, decor}, we adopt accuracy (\textbf{acc.}) and F1 score (\textbf{f1.}) to evaluate performance. For all experiments, we report the average of 5 independent runs.

\begin{table}[t]
\centering
\caption{Performance comparisons under our proposed temporality-aware setting.}
\vspace{-2ex}
\resizebox{0.8\linewidth}{!}{
\begin{tabular}{c l c c c c c c}
\hline
&\multirow{2}{*}{\textbf{Method}}&&\multicolumn{2}{c}{\textbf{PolitiFact}}&&\multicolumn{2}{c}{\textbf{GossipCop}}\\
\cline{4-5}
\cline{7-8}
&&&acc.&f1.&&acc.&f1.\\
\hline
\multirow{3}{*}{\textbf{G1}}&dEFEND{\textbackslash}c~\cite{defend}&&81.2&79.6&&75.4&68.3\\
&DualEmo{\textbackslash}c~\cite{dualemo}&&84.0&81.5&&77.9&71.9\\
&BERT~\cite{bert}&&84.2&81.7&&76.4&69.1\\
&GPT3.5&&75.7&77.9&&62.8&56.6\\
\hline
\multirow{6}{*}{\textbf{G2}}&GCNFN~\cite{gcnfn}&&84.8&82.3&&85.3&82.9\\
&UPFD~\cite{upfd}&&84.6&82.1&&82.0&77.8\\
&FANG~\cite{fang}&&85.5&82.6&&86.1&83.4\\
&GCN~\cite{gcn}&&86.4&83.0&&87.4&84.7\\
&GAT~\cite{gat}&&86.7&79.5&&86.5&83.6\\
&GraphSAGE~\cite{graphsage}&&85.4&80.9&&87.5&85.5\\
\hline
\multirow{2}{*}{\textbf{G3}}&RS-GNN~\cite{rsgnn}&&80.8&62.8&&85.9&83.6\\
&DECOR~\cite{decor}&&\underline{87.4}&\underline{84.7}&&\underline{88.1}&\underline{85.7}\\
\hline
\textbf{Ours}&\proposed&&\textbf{91.9}&\textbf{89.8}&&\textbf{93.7}&\textbf{93.0}\\
\hline
\end{tabular}
}
\label{table:main}
\vspace{-2ex}
\end{table}

\subsection{Detection Performance (RQ1)} \label{performance}
Table~\ref{table:main} compares the performance of \proposed~and baseline models, where the bold (underlined) values indicate the best (second best) results. We can observe that \textbf{(1)} social graph-based methods (G2) generally outperform content-based methods (G1), including LLMs. This highlights the importance of utilizing social contexts for fake news detection. \textbf{(2)} Within group G3, RS-GNN is shown to be less effective than DECOR. This indicates that GSL guided by node features is less suited for the fake news detection task as opposed to leveraging edge-specific features. \textbf{(3)} \proposed~ outperforms baseline models by a substantial margin, enhancing accuracy and F1 score by up to 5.6\%p and 7.3\%p over the most competitive baseline on the GossipCop dataset, respectively. This demonstrates that under the temporality-aware setting where accessible information and social structures change greatly before and after training,~\proposed~displays more robust detection performance despite its simpler methodology. As previously discussed, this is thanks to the proposed earliness-related patterns being based upon general social behaviors that occur similarly even at different points in time.

Additionally, we also investigate the performance of \proposed~and existing social graph-based methods when evaluated under the \textit{conventional temporality-ignorant setting}. In Fig.~\ref{fig:intro}(b), we observe that \proposed~ significantly outperforms baselines and displays marginal performance difference between the two settings as opposed to previous methods that suffer substantial degradation. Such results further highlight the effectiveness of our time-independent features and demonstrate the versatility of our proposed method, as it displays superior performance under various scenarios.

\begin{table}[t]
\centering
\caption{Ablation studies on~\proposed~(F1 score (\%)).}
\vspace{-2ex}
\resizebox{0.8\linewidth}{!}{
\begin{tabular}{c l|c c}
\hline
&&PolitiFact&GossipCop\\
\hline
\textbf{Ours}&\proposed&\textbf{89.8}&\textbf{93.0}\\
\hline
\multirow{5}{*}{\textbf{Feature Variants}}&\proposed+RAND&58.9&79.7\\
&\proposed-USER&85.6&92.5\\
&\proposed-ENG&83.2&92.1\\
&\proposed+RATIO&76.1&89.2\\
&\proposed+NF&72.0&80.7\\
\hline
\multirow{2}{*}{\textbf{Component Variants}}&\proposed-RANK&76.5&91.9\\
&\proposed+BC&62.4&90.2\\
\hline
\end{tabular}
}
\label{table:ablation}
\vspace{-4ex}
\end{table}

\subsection{Ablation Study (RQ2)} \label{ablation}
We conduct ablation studies to validate (1) the effectiveness of our earliness-related edge features proposed in Section~\ref{Joint Patterns}, and (2) the model components of~\proposed.
Specifically, \textbf{~\proposed+RAND} replaces our features with random values from the uniform distribution. \textbf{~\proposed-USER} and \textbf{~\proposed-ENG} remove user-wise and engagement-wise earliness patterns from the edge features, respectively. \textbf{~\proposed+RATIO} constructs edge features in an alternate way, by retrieving the relative ratio of each group (EE, EL, LE and LL) size. \textbf{~\proposed+NF} replaces our features with concatenated features of a node pair for each edge (i.e., concatenation of two 768-dimensional BERT embedding vectors). Meanwhile, \textbf{~\proposed-RANK} removes $\mathcal{L}_{rank}$ from the final loss function by setting $\alpha = 0$, and \textbf{~\proposed+BC} replaces $\mathcal{L}_{rank}$ with a binary classification loss as mentioned in Section~\ref{reweighter}.

The results are summarized in Table~\ref{table:ablation}. \proposed~ is shown to outperform all variants. In detail, \proposed+RATIO, ~\proposed+RAND and ~\proposed+NF display worse results when compared to ~\proposed, highlighting the effectiveness of our earliness-related edge features and supporting our observation in Section~\ref{performance}. Additionally, \proposed~ outperforms ~\proposed-USER and ~\proposed-ENG as well, which is in line with our analysis in Section~\ref{Joint Patterns} that jointly considering both user-wise and engagement-wise earliness patterns results in richer patterns than focusing on either one on its own.

Further, we observe that \proposed~ outperforms ~\proposed-RANK, proving the effectiveness of $\mathcal{L}_{rank}$. In addition, ~\proposed+BC is inferior to ~\proposed-RANK, supporting our claim that strictly sending edge weights to 0 or 1 is unfit for our task and hinders performance.

\begin{figure}[t]
\centering
\subfloat{
\includegraphics[width=0.95\linewidth]{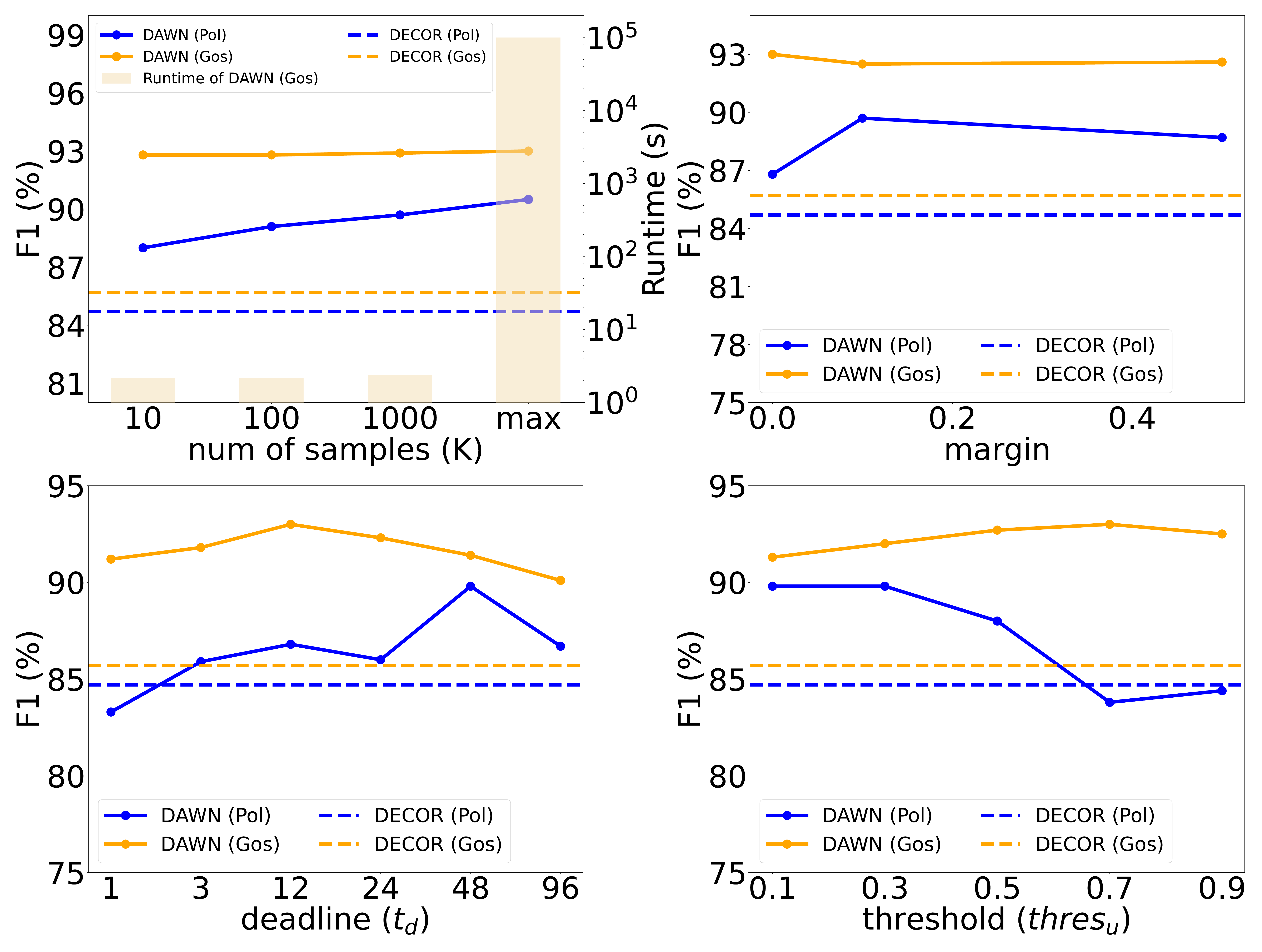}
}
\vspace{-2ex}
\caption{Hyperparameter sensitivity analysis.}
\vspace{-3ex}
\label{fig:hyper}
\end{figure}

\subsection{Hyperparameter Sensitivity (RQ3)} \label{hyperparameter}
In this section, we explore how the performance of \proposed~ is affected by varying the values of key hyperparameters, i.e., $K, margin$, $t_d$ and $thres_u$. The results are shown in Fig.~\ref{fig:hyper}.

We observe that \proposed~significantly outperforms DECOR under different variations of the sample size $K$ and $margin$ used in Equation~\ref{eqn:margin} (See Fig.~\ref{fig:hyper} top). The detection performance shows an upward trend when increasing $K$, which is expected since more edge pairs are involved during training\footnote{\textit{max} indicates comparing \textit{all} clean edge-noisy edge pairs instead of $K^2$ pairs.}. 
It is important to note that sampling only 10 edges per group (i.e., $K=10)$ significantly reduces computation time yet still shows performance comparable to the cases with larger $K$s, even surpassing DECOR, indicating the effectiveness of~\proposed\footnote{Bars in Fig.~\ref{fig:hyper} top denote log-scaled training time (s) of~\proposed~on GossipCop.}. 

\looseness=-1
In terms of deadline $t_d$ and threshold $thres_u$ (See Fig.~\ref{fig:hyper} bottom), we have some interesting observations. In detail, the performance on GossipCop peaks at a much shorter deadline $t_d$ than on PolitiFact, i.e., engagements need to occur much quicker to be considered early in GossipCop. Similarly, the performance on GossipCop is subpar when the threshold $thres_u$ is set to lower values, while on PolitiFact a threshold value too large significantly hinders performance, meaning early users respond much more quickly to news in Gossipcop. Both results indicate that what is considered ``early'' varies depending on the dataset, where much stricter standards should be used for GossipCop over PolitiFact. This can be attributed to the fact that readers tend to be much more attracted to celebrity gossip (i.e., GossipCop) than political news (i.e., PolitiFact)~\cite{preference, preference2}, resulting in stronger and quicker engagements. As such, applying stricter constraints when determining ``early'' user engagements can aid in distinguishing the earliness patterns of clean and noisy edges within GossipCop, while relatively lenient constraints are more appropriate for PolitiFact. Further investigating such trends for various news topics may provide valuable insights for future research.

\subsection{Efficiency on Large Networks (RQ4)} \label{eff}
To assess the applicability of \proposed~ on real-world tasks with massive social networks, we evaluate its computational cost on GossipCop as it is a much larger dataset compared with PolitiFact. Specifically, we train GCN, DECOR and \proposed~under identical settings, and report the mean runtime per epoch and resulting F1 score in Table~\ref{table:efficiency}. 
We observe that although \proposed~requires relatively higher computational cost, which is expected due to the introduction of $\mathcal{L}_{rank}$ in Equation~\ref{eqn:margin}, it achieves much quicker convergence, as shown in Fig.~\ref{fig:loss}. In practice, \proposed's fast convergence is likely to offset the slightly higher computational cost per epoch, leading to an overall more efficient training process. Further, we point out that the performance of \proposed~is significantly higher than that of GCN or DECOR. These results demonstrate \proposed's practical applicability in real-world fake news detection, as it substantially enhances performance while only modestly compromising efficiency.

\begin{table}[t]
\begin{minipage}{0.48\linewidth}{
    \centering
    \caption{Runtime and performance comparisons.}
    \vspace{-2ex}
    \resizebox{0.95\linewidth}{!}{
    \begin{tabular}{l c c c}
    \hline
    \multirow{2}{*}{\textbf{Method}}&&\multicolumn{2}{c}{\textbf{GossipCop}}\\
    \cline{3-4}
    &&Runtime (s)&f1.\\
    \hline
    GCN&&0.286&84.7\\
    DECOR&&0.544&85.7\\
    \proposed&&2.41&93.0\\
    \hline
    \end{tabular}
    }
    \label{table:efficiency}
}\end{minipage}
\begin{minipage}{0.48\linewidth}{
    \centering
    \subfloat{
    \includegraphics[width=0.9\linewidth]{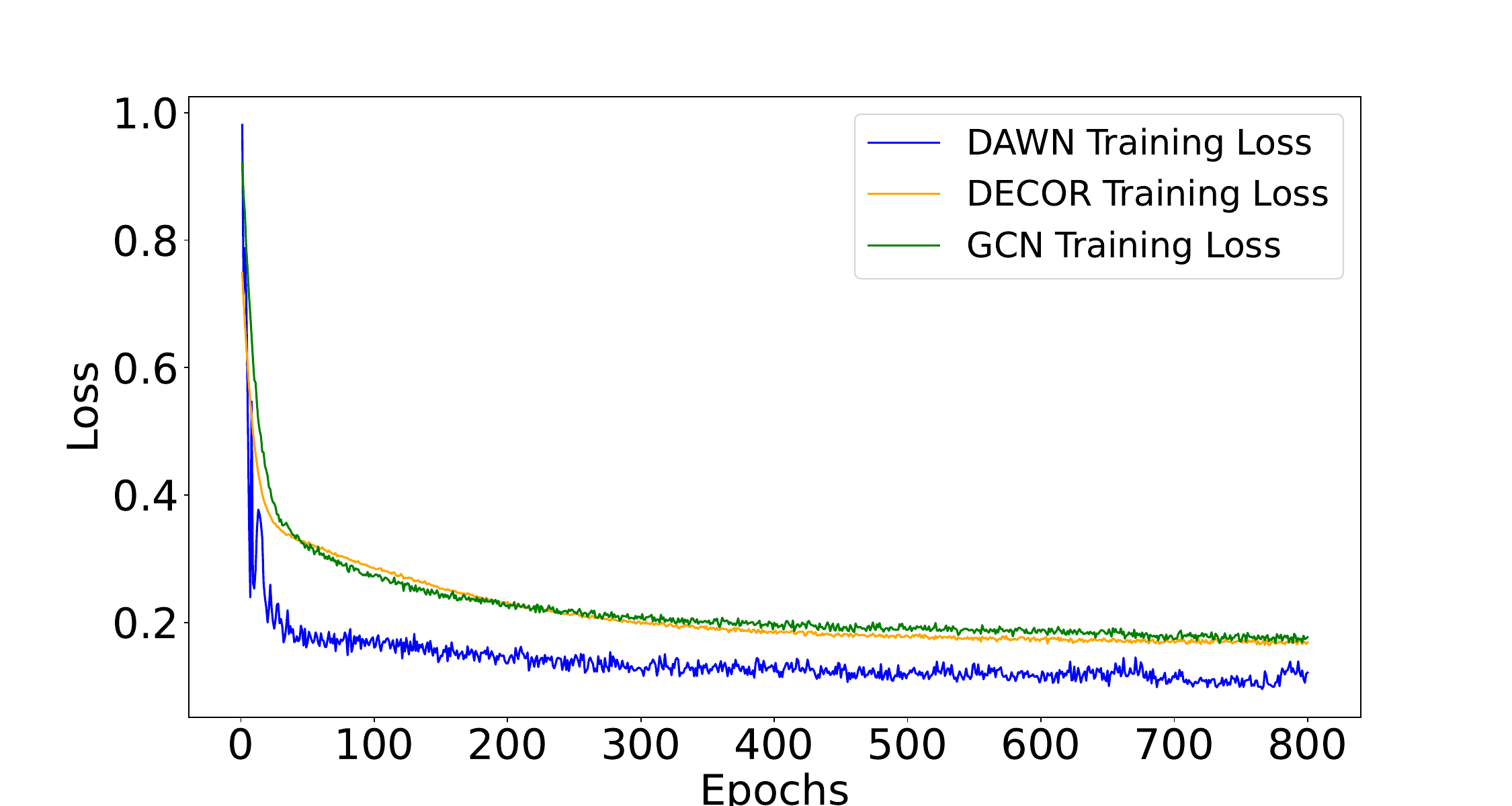}
    }
    \vspace{-5.5ex}
    \captionof{figure}{Model convergence comparisons.}
    \vspace{-1ex}
    \label{fig:loss}
}\end{minipage}
\vspace{-1ex}
\end{table}

\begin{table}[t]
\centering
\caption{Homophily ratio of social graphs before and after reweighting the edges by DECOR and ~\proposed.}
\vspace{-2ex}
\resizebox{0.95\linewidth}{!}{
\begin{tabular}{l|c c c}
\hline
&Original graph&Adjusted by DECOR&Adjusted by \proposed\\
\hline
PolitiFact&0.724&0.807&0.821\\
GossipCop&0.932&0.943&0.954\\
\hline
\end{tabular}
}
\label{table:homophily}
\vspace{-3ex}
\end{table}

\subsection{Case Study (RQ5)} \label{case}
We conduct a case study to further illustrate how \proposed~ successfully distinguishes and reweights clean and noisy edges, especially compared to a similar reweighting framework DECOR~\cite{decor}.

\begin{figure}[t]
\centering
\subfloat{
\includegraphics[width=0.9\linewidth]{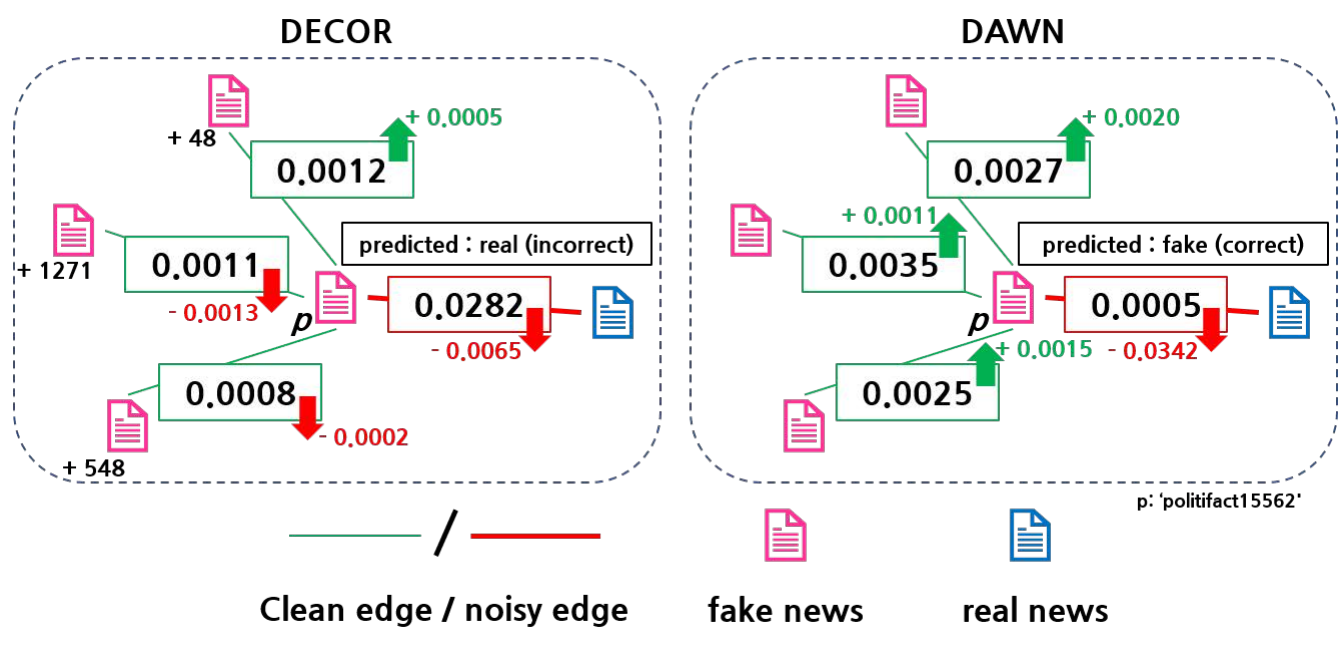}
}
\vspace{-2ex}
\caption{Case study comparison between DECOR (left) and \proposed~ (right). The boxed values denote the normalized edge weights adjusted by each method. The values next to the arrows show how much the weight has changed from the original adjacency matrix. 
For DECOR, the values next to the articles indicate the amount of degrees added in test time.
}
\vspace{-6ex}
\label{fig:case}
\end{figure}

In Fig.~\ref{fig:case} we compare the neighborhood of a fake news article $p$ after weight adjustment by DECOR (left) and \proposed~ (right). We observe that \proposed~ successfully upweights clean edges and significantly downweights the noisy edge, aiding in a correct prediction for $p$. On the other hand, DECOR wrongly downweights some clean edges and only marginally downweights the noisy edge, resulting in an incorrect prediction. Such differences are not limited to a single case, as we can see in Table~\ref{table:homophily} that \proposed~ exceeds DECOR in terms of the adjusted graph's homophily ratio (i.e., ($\sum$ clean edge weights) / ($\sum$ all edge weights)), proving \proposed~ is much more effective overall in suppressing noisy edges on the entire graph.

The reason for this is that DECOR's degree-based framework is unfit for temporality-aware settings. In detail, as test nodes and their related social contexts are unavailable during training, DECOR is vulnerable to unseen degree patterns that arise at test time. Further, even the degrees of training nodes are subject to substantial change at test time (See Fig.~\ref{fig:case} left, where nodes that experience a greater change in degree are more negatively impacted when adjusting their corresponding edge weights) when new related social contexts occur (e.g., reposting a very old news article). Both cases hinder DECOR's reweighting process, leading to performance degradation.

The earliness-related patterns utilized by \proposed, on the other hand, are based on \textit{earlier user engagements having a higher tendency to be drawn to articles with the same veracity}. As previously discussed, this simple yet universal social trend is deeply rooted in the well-explored confirmation bias theory, and thus occurs similarly even at different points in time. Due to the time-independent nature of our earliness features, \proposed~ can easily adapt to newly observed or substantially changed edges and adjust their weights accordingly, resulting in successful noisy edge suppression and robust effectiveness under the temporality-aware setting.

\section{Conclusion}
In this paper, we revisit the training and evaluation setting of social graph-based fake news detection methods, and present a more realistic temporality-aware setting where future data (both article-wise and context-wise) is unavailable during training. We further propose \proposed, a method more applicable to such scenarios that utilizes time-independent patterns rooted in fundamental social behaviors. Based on our observation that later user engagements contribute more to noisy edges that link real news-fake news pairs in the social graph, \proposed~ leverages feature representations of engagement earliness to suppress the weights of such noisy edges through a GSL framework. Extensive analysis on two prominent datasets show the robust effectiveness of our proposed method in detecting fake news under the temporality-aware setting, demonstrating the practical applicability of \proposed~ to real-world tasks.

\smallskip
\noindent{\textbf{Acknowledgements. }}
This work was supported by the National Research Foundation of Korea (NRF) grant funded by the Korea government (MSIT) (RS-2024-00335098, RS-2024-00406985), and by the Ministry of Science and ICT (NRF-2022M3J6A1063021).

\section*{Ethics Statement}
Regarding the adherence of ACM publication policy, to the best of our knowledge, there are no ethical issues with this paper. All datasets used for experiments are publicly available.

\bibliographystyle{ACM-Reference-Format}
\balance
\bibliography{reference}

\newpage
\appendix
\begin{algorithm}
\caption{Training Algorithm of \proposed}
\label{alg:main}
\begin{flushleft}
\textbf{Input:} $\mathcal{G}_{train} = (\mathcal{P}_{train}, \mathcal{A}_{train}), \mathcal{G}_{val} = (\mathcal{P}_{val}, \mathcal{A}_{val})$, $\mathcal{X}$, normalized edge features, $K, \alpha$ 

\textbf{Output:} Edge weight estimator $f$, GNN classifier $g$
\end{flushleft}
\begin{algorithmic}[1]
\STATE Randomly initialize the parameters of $f$ and $g$

\FOR{i = 1 to \# epochs}
    \STATE Get the reweighted training adjacency matrix $\mathcal{W}_{train}$ with $f$ by Equation~\ref{eqn:mlp}
    \STATE Input $\mathcal{W}_{train}$ and training node features from $\mathcal{X}$ to $g$ to get veracity predictions
    \STATE Randomly sample $K$ clean and noisy edges, respectively
    \STATE Jointly optimize parameters $\theta$ and $\phi$ by Equation~\ref{eqn:final}
    \STATE Get the reweighted validation adjacency matrix $\mathcal{W}_{val}$ and perform validation on $\mathcal{G}_{val}$
\ENDFOR

\STATE Return $f$ and $g$ best performing on $\mathcal{G}_{val}$

\end{algorithmic}
\end{algorithm}

\section{Training Algorithm of \proposed} \label{pseudocode}
We present the formal training process of \proposed~ in Algorithm~\ref{alg:main}. For the noisy edge suppression module, the time complexity is $\mathcal{O}(|\mathcal{A}_{train}| \cdot h_1 + K^2)$, where $|\mathcal{A}_{train}|$ denotes the number of existing edges in $\mathcal{A}_{train}$ and $h_1$ is the dimension size of the hidden layer of $f$. Note that $K << |\mathcal{A}_{train}|$. For the fake news detection module, the time complexity is $\mathcal{O}(|\mathcal{A}_{train}| \cdot h_2 + |\mathcal{P}_{train}| \cdot F \cdot h_2)$, where $h_2$ is the dimension size of the hidden layer of $g$, $|\mathcal{P}_{train}|$ denotes the number of training nodes and $F$ is the dimension size of node features. Thus, the overall time complexity of training \proposed~ is $\mathcal{O}(|\mathcal{A}_{train}| \cdot h_1 + K^2 + |\mathcal{A}_{train}| \cdot h_2 + |\mathcal{P}_{train}| \cdot F \cdot h_2)$.

Due to the addition of $\mathcal{L}_{rank}$ during training, ~\proposed~ requires a relatively higher computational cost compared to methods that only utilize prediction loss. In future work, we plan to mitigate this by introducing an optimization approach to ~\proposed, e.g., leveraging the Quicksort algorithm to make the learning process of the ranking loss more efficient~\cite{effrank}, thereby reducing the computational burden.

\section{Further Experimental Settings} \label{exp}
\subsection{Baselines}
\begin{itemize}[leftmargin = 2mm]
    \item \textbf{Content-based methods (G1)} utilize patterns found within the news article text. \textbf{dEFEND{\textbackslash}c} is a variant of dEFEND~\cite{defend} without considering user comments, which leverages a hierarchical attention framework. Similarly, \textbf{DualEmo{\textbackslash}c} is a variant of DualEmo~\cite{dualemo} without incorporating comment emotion, and concatenates both semantic and emotion features from the article text. We also compare with \textbf{BERT}~\cite{bert}, where the obtained article text embeddings are directly fed into MLP layers for prediction. Additionally, we report the performance of a representative LLM, i.e., \textbf{GPT3.5} (version name: gpt-3.5-turbo).

    \item \textbf{Social graph-based methods (G2)} additionally incorporate social contexts by constructing them into graph structures, and then learn article representations through GNNs. 
    Since~\proposed~relies solely on information from news articles, user identities and user-news engagements, we restricted the baselines to model components that utilize the same types of information for fair comparisons.
    \textbf{GCNFN}~\cite{gcnfn} and \textbf{UPFD}~\cite{upfd} construct propagation trees for each news article by utilizing user responses. \textbf{FANG}~\cite{fang} learns the representations of a heterogeneous social graph, consisting of users, news articles and sources. We additionally report the performance of three well-known GNN architectures, i.e., \textbf{GCN}~\cite{gcn}, \textbf{GAT}~\cite{gat} and \textbf{GraphSAGE}~\cite{graphsage}, on our social graph. Particularly, GraphSAGE is included due to its inductive scheme, as our temporality-aware setting results in inductive graphs.

    \item \textbf{Graph Structure Learning (GSL) methods (G3)} aim to downweight noisy edges within the graph by optimizing the adjacency matrix, thereby enhancing the learned representations. \textbf{RS-GNN}~\cite{rsgnn} utilizes a node feature similarity-guided link predictor, learning a denoised graph alongside a GNN. \textbf{DECOR}~\cite{decor} is a GSL method tailored to the fake news detection task, where degree-related edge features are fed into a link predictor to adjust edge weights. Note that while DECOR is also a social graph-based method, we placed it within this group for a clearer comparison.
\end{itemize}

\noindent{For all baselines, we follow the codes and hyperparameter values suggested by their respective authors. For GPT3.5, we set the temperature, top\_p and max\_token to 0.0, 1.0 and 256, respectively, and used the following prompt: ``Does the following contain fake news or information? Only answer with a single number; 1 if it contains fake news and 0 if it is real news. News article: [article text]''.}

\subsection{Implementation Details}
We execute \proposed~ based on PyTorch 1.12.0 with CUDA 12.4, and train them on a server supporting Rocky Linux 8.7 (Green Obsidian) with NVIDIA RTX A6000 GPU and Intel(R) Xeon(R) Gold 6326 CPU @ 2.90GHz. We set the number of engagements $m=3$ to obtain the set of active users for social graph construction. For the initial node features, we extract 768-dimensional embedding vectors from the article texts via a pre-trained BERT~\cite{bert} model with frozen parameters. The edge weight estimator $f$ is a 2-layer MLP with hidden size 16. The GNN classifier $g$ is a GCN~\cite{gcn} consisting of 2 layers with 64 hidden dimensions. \proposed~is trained for 1000 epochs, and the best model is selected through validation. For our main results, $(t_d, thres_u, \alpha, K, margin)$ are set to $(48hrs, 0.3, 0.1, 1000, 0.1)$ and $(12hrs, 0.7, 0.3, 10000, 0.0)$ for PolitiFact and GossipCop, respectively. All module parameters are optimized by the Adam optimizer~\cite{adam} with learning rate 0.001.

\section{Performance under Limited Training Data} 
\label{train_ratio}
\begin{figure}[t]
\centering
\subfloat{
\includegraphics[width=0.95\linewidth]{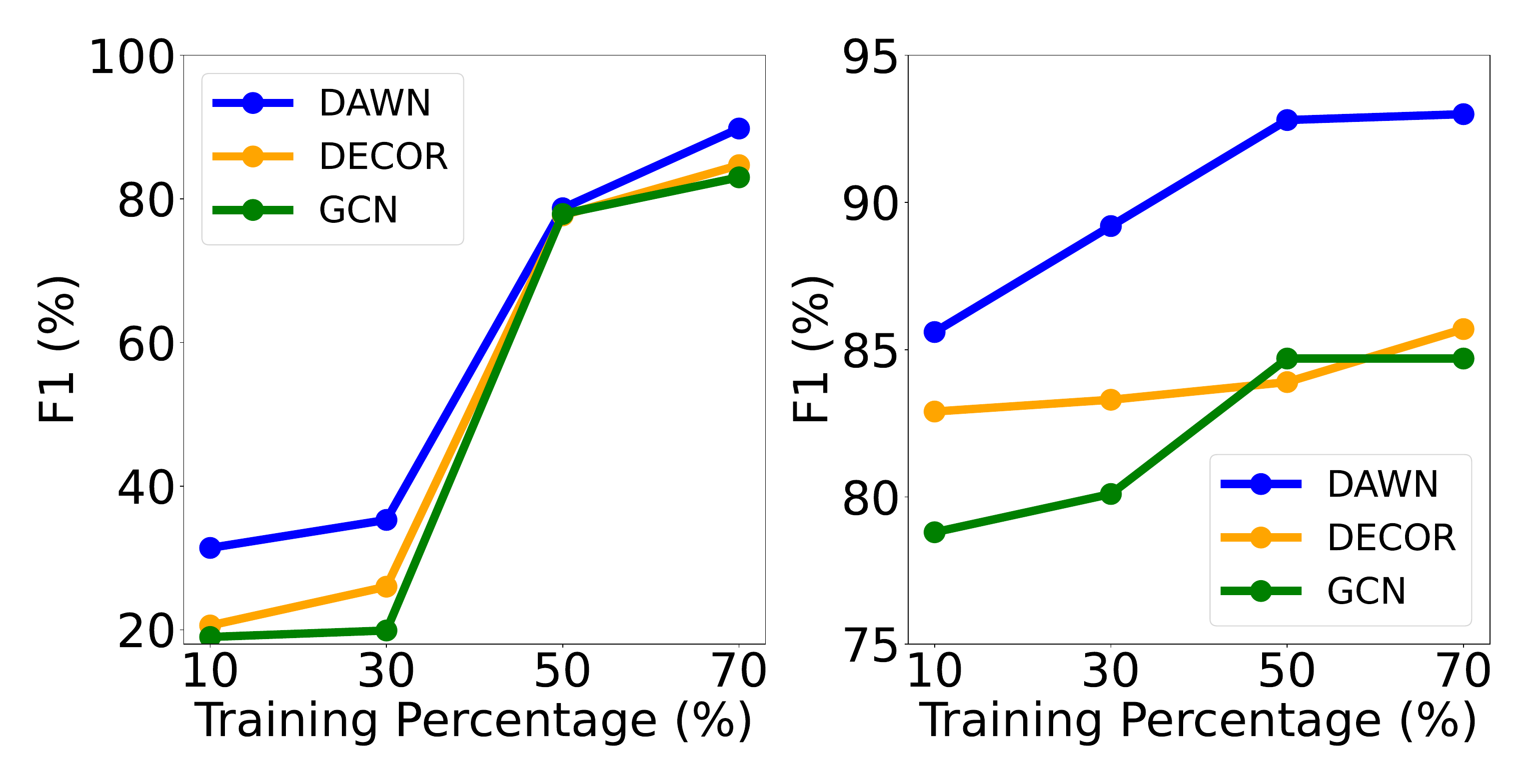}
}
\vspace{-2ex}
\caption{Performance comparisons between \proposed~ and baselines across various training data sizes.}
\vspace{-2ex}
\label{fig:train_ratio}
\end{figure}

To further assess the applicability of our proposed method to real-world fake news detection tasks, we evaluate the performance of \proposed~ under limited training data. In detail, we adjust the timestamps to allow varying training set size (i.e., 10\%, 30\%, 50\% and 70\% of news articles in temporal order). The social contexts contained in the training set would also change accordingly. We can observe in Fig.~\ref{fig:train_ratio} that \proposed~ consistently outperforms competitive baselines on both datasets. In particular, all models exhibit poor performance at training percentages 10\% and 30\% on PolitiFact, due to severe class imbalance present in the training set. Despite this, \proposed~ displays a significant performance gain over the baselines, demonstrating the effectiveness of our earliness-related features, i.e., our proposed method is able to leverage rich social patterns even under severely limited available data. These results indicate that \proposed~ is highly suitable for real-world applications, especially under the temporality-aware scenario where the collected data for training may be scarce.

\section{Generalizability of Earliness-Related Patterns}
In Section \ref{analysis}, our empirical analysis on two English fake news datasets revealed rich earliness-related patterns, indicating a closer relationship between earlier user engagements and veracity label consistency. In order to examine whether such patterns are limited to a single social media platform or specific topics and languages, we extended our analysis to the Multi-source Chinese Fake News Benchmark Dataset (MCFEND)~\cite{mcfend}, with descriptive statistics summarized in Table~\ref{table:stats_mcfend}.

\begin{figure}[t]
\centering
\subfloat{
\includegraphics[width=0.95\linewidth]{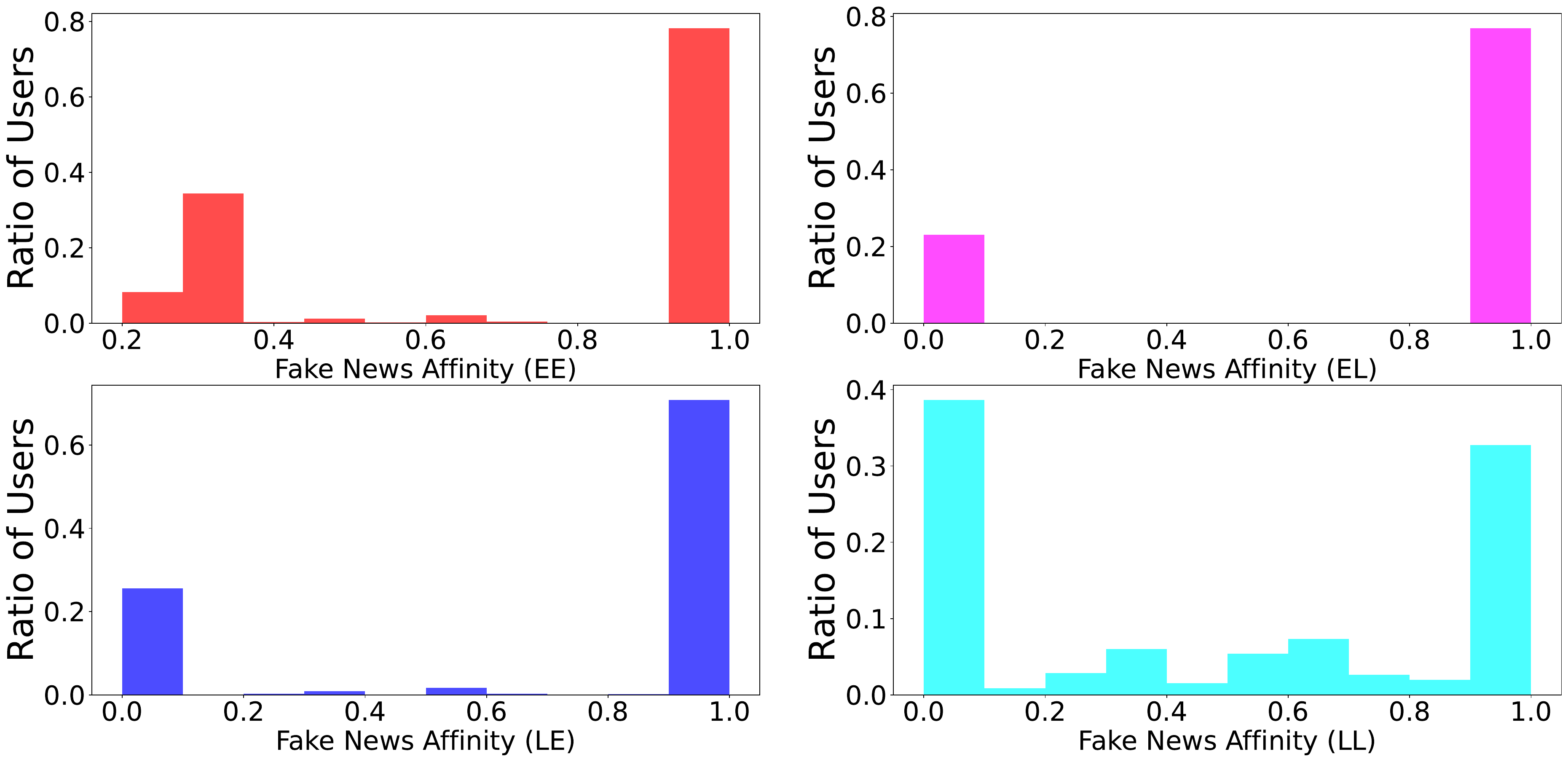}
}
\vspace{-2ex}
\caption{Joint earliness patterns for MCFEND.}
\vspace{-2ex}
\label{fig:joint_mcfend}
\end{figure}

Through the procedure detailed in Section \ref{Joint Patterns}, we plotted the FNA scores for MCFEND in Fig.~\ref{fig:joint_mcfend}, jointly taking into account both engagement-wise and user-wise perspectives ($t_d$ was set to one day, i.e., the minimum time unit in MCFEND, and $thres_u$ was set to 0.8). The resulting patterns are nearly identical to  what we observed in Fig.~\ref{fig:joint-plots}, where earlier user engagements tend to be more skewed towards either only fake news or real news. Such observations suggest that our earliness-related insights are applicable to broader contexts and not limited by the source, news topic or language; this in turn supports the generalizability of our proposed ~\proposed.

\begin{table}[t]
\centering
\caption{Dataset statistics for MCFEND.}
\vspace{-2ex}
\resizebox{0.5\linewidth}{!}{
\begin{tabular}{l c}
\hline
\textbf{Dataset}&\textbf{MCFEND}\\
\hline
\# News Articles&23,789\\
\# Real News&6,074\\
\# Fake News&17,715\\
\# Users&803,779\\
\# Engagements&2,102,902\\
\hline
\end{tabular}
}
\label{table:stats_mcfend}
\vspace{-3ex}
\end{table}

\end{document}